\documentclass[12pt]{article} \usepackage{amssymb,latexsym}
\usepackage{amsmath}
 
\columnwidth\textwidth

\begin{document}

\newtheorem{df}{Definition} \newtheorem{thm}{Theorem} \newtheorem{lem}{Lemma}
\newtheorem{rl}{Rule}
\begin{titlepage}
 
\noindent
 
\begin{center} {\LARGE Emergence of classical theories from quantum mechanics}
\vspace{1cm}

P. H\'{a}j\'{\i}\v{c}ek \\ Institute for Theoretical Physics \\ University of
Bern \\ Sidlerstrasse 5, CH-3012 Bern, Switzerland \\ hajicek@itp.unibe.ch

\vspace{1cm}

January 2012 \\

\vspace{1cm}
 
 
\vspace*{2cm}
 
\nopagebreak[4]
 
\begin{abstract} Three problems stand in the way of deriving classical
theories from quantum mechanics: those of realist interpretation, of classical
properties and of quantum measurement. Recently, we have identified some tacit
assumptions that lie at the roots of these problems. Thus, a realist
interpretation is hindered by the assumption that the only properties of
quantum systems are values of observables. If one simply postulates the
properties to be objective that are uniquely defined by preparation then all
difficulties disappear. As for classical properties, the wrong assumption is
that there are arbitrarily sharp classical trajectories. It turns out that
fuzzy classical trajectories can be obtained from quantum mechanics by taking
the limit of high entropy. Finally, standard quantum mechanics implies that
any registration on a quantum system is disturbed by all quantum systems of
the same kind existing somewhere in the universe. If one works out
systematically how quantum mechanics must be corrected so that there is no
such disturbance, one finds a new interpretation of von Neumann's "first
kind of dynamics", and so a new way to a solution of the quantum
measurement problem. The present paper gives a very short review of this work.
\end{abstract}

\end{center}

\end{titlepage}

\section{Introduction} All systems investigated by classical physics seem to
consist of quantum particles and their dependent fields. Then, they seem to be
also quantum systems and their classical properties ought to be derivable from
quantum mechanics. More precisely, we expect that one can find quantum models
of all classical properties. There are three independent problems that stand
in the way of a satisfactory completion of this projects.

First, classical systems and their properties are objective at least in the
sense that the assumption of their real existence, independent of
measurements, does not lead to any contradictions. Exactly this does not seem
to hold for quantum systems and their properties, which is called {\em Problem
of Realist Interpretation} of quantum mechanics. Second, classical properties
such as a position do not allow linear superposition. Nobody has ever seen a
table to be in a linear superposition of being simultaneously in the kitchen
as well as in the bedroom. Moreover, observing the table does not shift it
while quantum registration changes the state of the registered system. Let us
call this {\em Problem of Classical Properties}. Finally, evidence suggests
the assumption that the registration apparatus always is in a well-defined
classical state at the end of any quantum measurement indicating just one
value of the registered observable. This is called {\em objectification
requirement} \cite{BLM}. However, if the registered system is in a linear
superposition of different eigenvectors of the observable, then the linearity
of Schr\"odinger equation implies that the apparatus is also in a linear
superposition eigenstates of its pointer observable. Let us call this {\em
Problem of Quantum Measurement}.

Our work during the recent five years has lead to a surprising discovery: The
problems are considerably aggravated by some inveterate tacit assumptions
about quantum mechanics that are clearly wrong if one states them explicitly
\cite{HT,hajicek1, hajicek2,survey,hajicek3,hajicek4}. Quantum theory purified
from these errors has been called {\em The Reformed Quantum Mechanics}. The
present paper gives a very short review of our main ideas. The paper starts
with the Problem of Quantum Measurement because only this requires changes in
the conceptual and mathematical structure of quantum mechanics, proceeds to
the Problem of Realist Interpretation and finishes by the Problem of Classical
Properties.

\section{Problem of Quantum Measurement} In classical physics, any measurement
can be done, at least in principle, so that the system on which we measure is
not disturbed by the measurement. We assume, as did the co-called Copenhagen
interpretation, that this is not true for quantum measurement. A {\em
registration} of any observable on quantum system ${\mathcal S}$ needs a
special device ${\mathcal A}$, the so-called {\em registration apparatus}, and
the registration is always an interaction between ${\mathcal S}$ and
${\mathcal A}$ that changes the state of the ${\mathcal S}$, except for some
special cases. Moreover, to speak of the state of ${\mathcal S}$, there must
be a {\em preparation}. Thus each measurement on microsystems can be split
into preparation and registration.

The errors that cause the confusion in the theory of quantum measurement are
two: 1) neglecting the disturbance of registration due to identical particles
and 2) disregarding the role and structure of detectors.

\subsection{Disturbance of registration due to identical particles} Let us
consider two experiments.
\par \vspace*{.4cm} \noindent {\bf Experiment I}: State $\psi(\vec{x}_1)$ of
particle ${\mathcal S}_1$ is prepared in our laboratory.
\par \vspace*{.4cm} \noindent {\bf Experiment II}: State $\psi(\vec{x}_1)$ is
prepared as in Experiment I and state $\phi(\vec{x}_2)$ of particle ${\mathcal
S}_2$ of the same type is prepared simultaneously in a remote laboratory.
\par \vspace*{.4cm} \noindent If our laboratory does not know about the second
one, it believes that the state of ${\mathcal S}_1$ is $\psi(\vec{x}_1)$. If
it does then it believes that the state is
\begin{equation}\label{symstate} 2^{-1/2}\bigl(\psi(\vec{x}_1)\phi(\vec{x}_2)
\pm \phi(\vec{x}_1)\psi(\vec{x}_2)\bigr)\ .
\end{equation}

Thus, it seems that the notions of preparation and of state are ambiguous. Has
this ambiguity any observable consequences? To answer this question, let us
first consider Experiment I supplemented by a registration corresponding to
the observable of ${\mathcal S}_1$ with kernel $a(\vec{x}_1;\vec{x}'_1)$, and
let the registration be made in our laboratory. Measurements of this kind lead
to average value
\begin{equation}\label{aver1} \int d^3x_1d^3x'_1\,
a(\vec{x}_1;\vec{x}'_1)\psi^*(\vec{x}_1)\psi(\vec{x}'_1)\ .
\end{equation} Second, perform Experiment II supplemented by the registration
by the same apparatus in our laboratory as above. Because the apparatus cannot
distinguish between the contributions by the two particles, the correct
observable corresponding to this registration now is:
\begin{equation}\label{symobs}
a(\vec{x}_1;\vec{x}'_1)\delta(\vec{x}_2-\vec{x}'_2) +
\delta(\vec{x}_1-\vec{x}'_1)a(\vec{x}_2;\vec{x}'_2)\ .
\end{equation} Such measurements lead to the average value defined by Eqs.\
(\ref{symstate}) and (\ref{symobs}):
\begin{equation}\label{aver2} \int d^3x_1d^3x'_1\,
a(\vec{x}_1;\vec{x}'_1)\psi^*(\vec{x}_1)\psi(\vec{x}'_1) + \int
d^3x_1d^3x'_1\, a(\vec{x}_1;\vec{x}'_1) \phi^*(\vec{x}_1)\phi(\vec{x}'_1)\ .
\end{equation} Expression (\ref{aver2}) appreciably differs from (\ref{aver1})
for all standard observables such as position, momentum, spin, angular
momentum energy etc.

We conclude that the registrations of the standard quantum observables on a
system is always disturbed by all other systems of the same kind existing
somewhere in the universe.

\subsection{Separation status} Cluster separability principle is a kind of
locality assumption that has been fruitful in several branches of quantum
physics, see e.g.\ \cite{KP} and Chap.\ 4 of \cite{weinberg}. An application
of the principle to identical particles can be found for instance in
\cite{peres} p.\ 128. We formulate it as follows:
\par \vspace*{.4cm} \noindent {\bf Cluster Separability Principle} {\itshape
No quantum experiment with a system in a local laboratory is affected by the
mere presence of an identical system in remote parts of the universe.}
\par \vspace*{.4cm} \noindent This is violated by results similar to those of
the previous section but confirmed by experience. We must therefore correct
quantum mechanics so that it satisfies the Cluster Separability Principle.

We introduce an important locality property of observables \cite{hajicek2}
(for generalisation to composite systems, see \cite{survey}; a similar local
condition on observables has been introduced in \cite{wan,wanb}):
\begin{df} Let $D \subset {\mathbb R}^{3}$ be open. Operator with kernel
$a(\vec{x}_1;\vec{x}'_1)$ is \underline{$D$-local} if
$$
\int d^3x'_1\, a(\vec{x}_1;\vec{x}'_1) f(\vec{x}'_1) = \int d^3x_1\,
a(\vec{x}_1;\vec{x}'_1) f(\vec{x}_1) = 0\ ,
$$
for any test function $f$ that vanishes in $D$.
\end{df} Now assume for Experiment II, to keep everything simple, that
\begin{enumerate}
\item our laboratory is inside open set $D \subset {\mathbb R}^{3}$,
\item $\text{supp}\,\phi \cap D = \emptyset$.
\end{enumerate} Then, the second term in (\ref{aver2}) vanishes for all
$D$-local observables and Eqs.\ (\ref{aver1}) and (\ref{aver2}) agree in this
case (for a more general theorem see \cite{survey}). This suggests the
following approach.
\begin{enumerate}
\item We introduce the key notion of our theory:
\begin{df} Let ${\mathcal S}$ be a particle and $D \subset {\mathbb R}^3$ an
open set satisfying the conditions:
\begin{itemize}
\item Registrations of any $D$-local observable ${\mathsf A}$ of ${\mathcal
S}$ lead to average $\langle \psi(\vec{x})|{\mathsf A}\psi(\vec{x})\rangle$
for all states $\psi(\vec{x})$ of ${\mathcal S}$.
\item ${\mathcal S}$ is prepared in state $\phi(\vec{x})$ such that
supp\,$\phi \cap D \neq \emptyset$.
\end{itemize} In such a case, we say that ${\mathcal S}$ has
\underline{separation status} $D$.
\end{df} For generalisation to composite systems and non-vector states, see
\cite{survey}.\footnote{Further generalisation is possible. We are working on
two: First, to replace the condition supp\,$\phi \cap D \neq \emptyset$ by
$\int_D d^3x |\phi|^2 < \epsilon$ and that on the registered averages to the
averages lying in intervals $(\langle \psi(\vec{x})|{\mathsf
A}\psi(\vec{x})\rangle - \epsilon',\langle \psi(\vec{x})|{\mathsf
A}\psi(\vec{x})\rangle + \epsilon')$, the epsilons being related to the finite
sensitivity of registration apparatuses. Second, to impose conditions on the
ranges of momenta defined by the sensitivity of detectors in addition to those
on coordinates.}
\item We assume: Any preparation of ${\mathcal S}$ must give it a non-trivial
separation status $D \neq \emptyset$. Then $D$-local observables are
individually registrable on ${\mathcal S}$ but \underline{only these are}.
\end{enumerate}

\subsection{Registration} Preparation transfers a trivial into a non-trivial
separation status. Thus, the separation status changes during a
preparation. What is the relation of registrations to separation status
change?

An important assumption of our theory of measurement is:
\par\vspace*{.4cm}\noindent {\bf Pointer hypothesis} {\itshape Any
registration apparatus for \underline{microsystems} must contain at least one
detector and every 'reading of a pointer value' (see e.g.\ \cite{BLM}) is a
macroscopic signal from a detector.}
\par\vspace*{.4cm}\noindent We take the definition of detector from
experimentalists \cite{leo,stefan}. Then, during registration, the system must
enter the sensitive matter of a detector and its non-trivial separation status
changes into a trivial one \cite{hajicek2,survey}.

In this way, preparation and registration have even more importance in our
theory than in Copenhagen interpretation: they must include changes of
separation status. Clearly, any change of separation status of a system
${\mathcal S}$ is also a change of its observable algebra, i.e., change of the
kinematics of ${\mathcal S}$ \cite{hajicek3}.

What has been said up to now shows that textbook quantum mechanics is
incomplete in the following sense:
\begin{enumerate}
\item It accepts and knows only \underline{two} separation statuses:
\begin{enumerate}
\item that of isolated systems, $D = {\mathbb R}^3$, with all self-adjoint
operators as observables, and
\item that of a member of a system of identical particles, $D = \emptyset$,
with no individual observables of its own.
\end{enumerate}
\item It provides \underline{no} rules for changes of separation status.
\end{enumerate} Hence, quantum mechanics must be supplemented by a theory of
general separation status \cite{hajicek2}, a correct theory of observables
\cite{hajicek2} and by new rules that govern processes in which separation
status changes \cite{hajicek4}.

In \cite{hajicek4}, a general rule for interaction between microsystem
${\mathcal S}$ and macrosystem ${\mathcal A}$ has been motivated and
discussed. Formally, the process is decomposed into three steps. First, there
is a change of kinematics and second, the unitary evolution of the state of
${\mathcal S} + {\mathcal A}$ in the framework of the new kinematics. Finally,
the state of ${\mathcal S} + {\mathcal A}$ given by the unitary evolution must
be corrected by a state reduction at the time of any detector signal in a way
that is uniquely determined by the signals. The decomposition into the three
steps is not the description of how things proceed in time: the rule gives
only the end state.

The state reduction itself consists of two changes. First, some non-diagonal
elements of the state operator of the total isolated system are
erased. Second, the resulting mixture is postulated to be a proper one (a
gemenge, see \cite{hajicek2}). Something similar to the first change can also
be achieved for an open subsystem of the total isolated system within standard
quantum mechanics by methods such as quantum decoherence \cite{BLM}. However,
the second change can only be accomplished (or circumvented) with help of new
assumptions that must be added to quantum mechanics such as the Everett
interpretation in the case of decoherence, see \cite{BLM,BG}. For a discussion
of super-selection methods such as Wan's \cite{wan}, see \cite{hajicek4}.

The Reformed Quantum Mechanics thus returns to von Neumann's "two kinds
of dynamics" (see also \cite{hajicek2}) but its notion of state reduction
differs from von Neumann's in two points. First, it is less ad hoc because it
is justified by the argument of separation status change, which is logically
independent from the proper quantum measurement problem and, second, it is
more specific because it happens only in a detector and its form is determined
by objective processes inside the detector sensitive matter including approach
to thermal equilibrium.

\section{The problem of realist interpretation} The very subject of quantum
mechanics is usually formulated very cautiously, e.g., \cite{peres}, p.\ 13:
\begin{quote} \dots quantum theory is a set of rules allowing the computation
of {\em probabilities} for the outcomes of tests [registrations] which follow
specific preparations.
\end{quote} This can be contrasted with naive expectation that
\begin{quote} quantum theory studies properties of existing microscopic
objects.
\end{quote} The cause of such reserve is an apparent lack of objective
properties of quantum systems. For example, one cannot ascribe values of
observables to quantum systems prior to registrations, or else there will be
nasty contradictions such as those due to contextuality \cite{peres}, p.\
187. An existing object must have enough objective properties. Thus, the very
existence of quantum systems is uncertain, see, e.g., \cite{peres}, p.\ 13.

The origin of this difficulty is the surprisingly inveterate tacit assumption
that the values of observables of a quantum system ${\mathcal S}$ are a kind
of basic properties of ${\mathcal S}$ so that no other properties can be
accepted in their place.

In our opinion, every value $a$ of observable ${\mathsf A}$ is only created
during registration of ${\mathsf A}$ by an apparatus ${\mathcal A}$ and it is
an objective property of composite ${\mathcal S} + {\mathcal A}$, not
${\mathcal S}$. It contains only some indirect information about ${\mathcal
S}$. Then, all the enormous effort to justify values of observables of
${\mathcal A}$ as properties of ${\mathcal S}$ is futile and must be
rejected. Instead, we postulate \cite{HT,survey}:
\par\vspace*{.4cm}\noindent {\bf Basic Ontological Hypothesis of Quantum
Mechanics} {\itshape A property is objective if its value is uniquely
determined by a preparation according to the rules of quantum mechanics. The
"value" is the value of the mathematical expression that describes
the property and it may be more general than just a real number. No
registration is necessary to establish such a property but correct
registrations cannot disprove its value; in most cases, registrations can
confirm the value.}
\par\vspace*{.4cm}\noindent Thus, we associate objective properties with
preparations instead of registrations. Let us give some obvious examples of
objective properties. First, the so-called {\em structural properties}: mass,
charge, spin, structure of Hamiltonian for isolated systems, etc. Second, the
so-called {\em dynamical properties}: state operator (not just wave function),
average value and variance of an observable, etc.

It may be helpful to recall that preparations include more general processes
than just manipulations in a human laboratory. For example, solar neutrinos
are prepared at the centre of the Sun. But even if a property is defined by
such a manipulation, the assumption that it can be ascribed to the prepared
system is non-trivial and helpful. In fact, this is also assumed in classical
physics for systems created in laboratory. In any case, a preparation is
defined by objective (classical) conditions on ${\mathcal S}$, which determine
the statistics of subsequent registrations on ${\mathcal S}$. More about the
relation between objective and subjective aspects of probability theory can be
found in \cite{survey}.

We have shown in \cite{survey} that all properties determined by preparations
form a Boolean lattice similarly to all objective properties of a classical
system. Hence, there is no need for an extra logic in quantum mechanics.

Moreover, one can completely describe the dynamical situation of any quantum
system by these objective properties. Hence, the condition on a system to be a
physical object are satisfied. Such an object---a real quantum system---is
theoretically modelled by an {\em occupied quantum state} \cite{survey}. In
all considerations about quantum systems, it is then possible to refer
directly to their objective properties, e.g., to speak of registration devices
instead of observers. "Unspeakable" becomes "speakable" in
quantum mechanics and it is a surprising liberation.

\section{The problem of classical properties} Quantum mechanics leads to
linear superposition of different states of macroscopic systems but classical
theories such as Newtonian mechanics and Maxwellian electrodynamics do not
allow such superpositions. How can this feature of the theories be derived
from quantum mechanics?

The origin of all difficulties to obtain such derivation is a common
inveterate assumption that a sharp trajectory is the key objective property of
classical systems. Hence, quantum models of classical systems usually start
from states with minimum uncertainty. These are pure and can easily be
linearly superposed.

However, a completely sharp classical trajectory is a figment of
imagination. {\em Real} states of all classical theories always have a
non-zero variance, which is always much larger than the minimum quantum
uncertainty. This is an old idea, cf.\ \cite{exner,born}. It is, therefore,
sufficient that quantum models of classical systems explain such fuzzy
states. Instead of sharp trajectories, we assume:
\par\vspace*{.4cm}\noindent {\bf Basic classicality hypothesis} {\itshape To
construct models of classical systems, high-entropy states of macroscopic
quantum systems must be used.}
\par\vspace*{.4cm}\noindent This seems to be in accord with state reduction
being associated with approach to the thermal equilibrium of a detector (Sec.\
2.3). Moreover, high-entropy states are mixed and cannot be linearly
superposed. Of course, it is possible (but difficult) to prepare other states
of macroscopic systems: strong laser beams, large EBC, etc. This simply means
that not every state of a macroscopic system must be classical.

An example of suitable states modelling Newton mechanics of mass centre are
the so-called {\em maximum entropy (ME) packets}: they maximise entropy for
fixed averages and variances of coordinates and momenta \cite{hajicek1}:
\begin{thm} Let ${\mathcal S}$ be a quantum system with coordinates ${\mathsf
q}_k$ and momenta ${\mathsf p}_k$, $k=1,\cdots,n$. Then the state operator of
the ME packet with averages and variances $Q_k$, $\Delta Q_k$ of coordinates
and $P_k$, $\Delta P_k$ of momenta is
\begin{equation}\label{32.1} {\mathsf T}(Q,P,\Delta Q,\Delta P)=
\prod_{k=1}^n\left[\frac{2}{\nu_k^2-1}\exp\left(-\frac{1}{\hbar}\ln\frac{\nu_k+1}{\nu_k-1}\
{\mathsf K}_k\right)\right]\ ,
\end{equation} where
\begin{equation}\label{20.3} {\mathsf K}_k = \frac{1}{2}\frac{\Delta
P_k}{\Delta Q_k}({\mathsf q}_k-Q_k)^2 + \frac{1}{2}\frac{\Delta Q_k}{\Delta
P_k}({\mathsf p}_k-P_k)^2
\end{equation} and
\begin{equation}\label{20.3b} \nu_k = \frac{2\Delta P_k\Delta Q_k}{\hbar} \in
(1,\infty)\ .
\end{equation}
\end{thm} Here, ${\mathsf K}_k$ is {\em not} the Hamiltonian of ${\mathcal
S}$: the Hamiltonian can be arbitrary. A similar theorem can be shown for
classical EM packets \cite{hajicek1}.

Von Neumann entropy of ${\mathsf T}(Q,P,\Delta Q,\Delta P)$ is
\begin{multline*} S({\mathsf T}(Q,P,\Delta Q,\Delta P)) = tr[{\mathsf
T}(Q,P,\Delta Q,\Delta P)\ln{\mathsf T}(Q,P,\Delta Q,\Delta P)] = \\
\sum_{k=1}^n\left(\frac{\nu_k +1}{2}\ln(\nu_k +1) - \frac{\nu_k
-1}{2}\ln(\nu_k -1) - \ln 2 \right)
\end{multline*} and can be shown to be an increasing function of
$\nu_k$. Limit
$$
\nu_k \rightarrow 1
$$
for all $k$ gives
$$
S({\mathsf T}(Q,P,\Delta Q,\Delta P)) \rightarrow 0\ ,\quad {\mathsf
T}(Q,P,\Delta Q,\Delta P) \rightarrow |Q,P,\Delta Q,\Delta P\rangle \langle
Q,P,\Delta Q,\Delta P|\ ,
$$ 
where $|Q,P,\Delta Q,\Delta P\rangle$ is the Gaussian wave packet for the
averages and variances $Q,P,\Delta Q,\Delta P$.

The time dependence of the averages and variances can be calculated from ME
packets as initial states using the Hamiltonian of ${\mathcal S}$ in both
Newtonian and quantum mechanics. Comparison of functions ${\mathsf Q}_k(t)$,
${\mathsf P}_k(t)$, $\Delta{\mathsf Q}_k(t)$ and $\Delta{\mathsf P}_k(t)$
obtained in the two theories shows full agreement for all Hamiltonians
(potential functions) if
$$ 
\nu_k \rightarrow \infty
$$
for all $k$. Thus, classical limit is just the opposite of the textbook one:
it is not the Gaussian wave packet but a high entropy limit.

Observe that classical properties ${\mathsf Q}_k(t)$, ${\mathsf P}_k(t)$,
$\Delta{\mathsf Q}_k(t)$ and $\Delta{\mathsf P}_k(t)$ are uniquely determined
by the preparation process and are therefore objective. More discussion can be
found in \cite{HT} and \cite{survey}.

We close this section by a few words on semi-classical (or WKB)
approximation. It is an approximation method within quantum mechanics, usually
defined as the limit $\hbar \rightarrow 0$ in some quantum expressions
\cite{peres}. The resulting equations may be similar to the corresponding
classical equations. Limit $\nu \rightarrow \infty$ also results from $\hbar
\rightarrow 0$, if the variances are kept constant. However, the
semi-classical approximation alone is not sufficient for a derivation of
complete classical theories \cite{peres}.

\section{Conclusion}
\begin{enumerate}
\item Quantum mechanics is as objective as any other theory of physics.
\item Classical objects can be modelled by high-entropy states of macroscopic
quantum systems. Classical limit is a suitably taken high-entropy
limit. Coherent states have nothing to do with classical limit.
\item The theory of observables and of quantum measurement must be corrected
to remove the disturbance due to identical particles. This leads to a solution
of quantum measurement problem based on a state reduction. Our state
reduction, however, is less ad hoc than von Neumann's because it is justified
by a change of separation status, and it is more specific because it is
associated with objective processes in the sensitive matter of a detector.
\end{enumerate}

\subsection*{Acknowledgements} The author is indebted to Ji\v{r}\'{\i} Tolar
for discussion.

\end{document}